\documentclass[preprint,11pt]{elsarticle}

\usepackage{amssymb}
\usepackage{amsmath}
\usepackage{listings}%
\usepackage[version=4]{mhchem}
\usepackage{siunitx}
\usepackage{xurl}

\journal{Nuclear Physics B}

\begin{document}

\begin{frontmatter}



\title{On a variational model for phase transformation in \ce{SiO2} glass} 


\author{Sarah Dinkelacker-Steinhoff, Klaus Hackl} 

\affiliation{organization={Chair of Mechanics - Materials Theory, Ruhr University Bochum},
            addressline={Universit\"atsstrasse 150}, 
            city={Bochum},
            postcode={44801}, 
            state={North Rhine-Westphalia},
            country={Germany}}

\begin{abstract}
The compaction mechanisms of \ce{SiO2} glass under pressure include under certain conditions a specific reduction of the elastic moduli and a complex inelastic behavior whose nature is not yet fully understood. \\
In our work we establish a variational framework describing the evolution of \ce{SiO2} glass under hydrostatic pressure. Based on a previous work that presents a model for multi-phase transformations in inelastic materials, we assume isothermal conditions during a compaction process and interpret the typical sigmoidal stress response as indicator of a binary phase transformation.\\
During the process, two volume fractions coexist macroscopically and microstructures such as shear bands or disclination pattern develop in between. We restrict our approach to resolve only the volume fractions, not the corresponding microstructures. Nevertheless, the resulting model is shown to match experimental findings very well. 
Numerical examples successfully illustrate the relationship between the changes in the elastic moduli and the corresponding change in volume with respect to pressure.
\end{abstract}


\end{frontmatter}


\section{Introduction}\label{sec1}

Amorphous metal oxides such as \ce{GeO2} or \ce{SiO2} show a so-called anomalous behavior of their elastic moduli \cite{Mysen,Trachenko}. During compression, these moduli initially decrease to a minimum value with increasing pressure. This softening effect has been studied more frequently in recent years. As a result, for example, silica glass exhibits plastic behavior and reversible and non-reversible deformations have been observed under various experimental conditions. In contrast to earlier measurements \cite{Tsiok}, there is nowadays the ability to realize a pseudo-hydrostatic state over Diamond Anvil Cells (DAC) with an appropriate pressure medium \cite{Deschamps14}. Data and methods of measurements of glasses are already available for geological pressure scales from 0 GPa up to 130 GPa \cite{Petitgirard}. For silica glass, two main regions of structural change can be identified. In the interval from 0 GPa to approx. 20 GPa, no modification was confirmed for the \ce{SiO2} coordination, which is 4-fold at room temperature and ambient pressure. However changes occur in the so-called intermediate region, which ranges across structural units of (10–20) Å ~\cite{Du}. In order to better understand the anomalous behavior in the low pressure range, reference is often made to the so-called free volume theory.
Pure \ce{SiO2} glass has a higher free volume than glass structures modified with network modifiers \cite{Molnar, Huang1}. In this context, the term network modifiers refers, in a technical sense, to oxides that exhibit less tendency toward network-forming behavior. This is related to the chemical bonding behavior and the bonding energies, see Liebau \cite{Liebau} for more details. \\ Furthermore, phase transformations of their crystalline equivalents are often used for glasses to approximate the actual structure change. In the case of \ce{SiO2} glass, $\alpha$ - quartz with its wide-meshed tetrahedron network and its polymorphs, serves as comparison \cite{Mysen,Pabst,Huang1}. The transition between $\alpha$- and $\beta$ cristobalit is characterized in Huang et al. (2003) with the help of molecular dynamic (MD) simulations \cite{Huang}. This transition includes features which can be associated with the amorphous-amorphous transitions in silica glass, which effects changes at medium range. Besides this change between six-membered rings with lower and higher symmetry also the ring size distribution has to be considered, which is centered on a six-membered structure that changes with increasing compression \cite{Davila}.\\
Above 21 GPa, the bonding changes from 4-fold to 6-fold or higher. Whereas a 5-fold  structure is still being discussed \cite{Petitgirard}. In summary, the different measurements under hydrostatic pressure show, that the elastic limit is reached at 8 - 9 GPa \cite{Deschamps09, Bruns,Keryvin}. The function of the bulk modulus against pressure reaches a minimum at approx. 2 - 3  GPa \cite{Deschamps09,Deschamps14, Huang1,Mysen, Tsiok}. \\
If only the first pressure interval is considered, the density change is an important intensive factor.
According to Mysen \cite{Mysen}, it is difficult to accurately determine the density of glass in high-pressure experiments beyond the elastic limit. At room temperature, the density of fused silica is approx. $\rho_0$ = 2.20 \si{\g\per\cm^3} \cite{Pabst}, while it increases to approx. $\rho_{sat}$ = 2.60 \si{\g\per\cm^3} at a maximum pressure between 18 - 20 GPa or 25 GPa with microstructural changes \cite{Sonneville,Deschamps14}. The resulting relative density change $ \frac{\Delta \rho}{\rho_0}$  develops up to 21\%. If it is linked to the hydrostatic pressure, the equation of state (EOS) can be described by a sigmoidal shape. This relation, which can be seen as constitutive equation, is often given by evaluation of Brillouin scattering or Raman spectra \cite{Bruns,Rouxel,Deschamps14,Li,Sonneville}. \\
From the macroscopic point of view, one explanation for this shape could be, that there is a phase transformation between to volume fractions. In between, two phases may coexists and microstructure develops \cite{Schill1}, as it is also the case in other amorphous materials such as metallic glasses. Furthermore, these microstructures are in turn comparable to the optically visible shear bands that occur in soils. In contrast to amorphous metal oxides or bulk metallic glasses, it is already known that shear bands in soils are caused by an increased volume and a lower density in local zones. This dilatancy effects are linked to plastic flow \cite{Wolf}.\\
Various continuum mechanical models provide access to the behavior of silica glass under compression \cite{Li, Schill1, Keryvin}. Most of them are developed using experimental data or data created by MD simulations. The material model is then formulated using incorporated fitting functions. In addition, it is known, that the supporting three-dimensional MD-simulations show data that deviate from the high-pressure experiments \cite{Huang1}. The data created using MD-simulations displaying higher or lower values. Furthermore, little attention is often paid to the relationship between the resulting curves.\\
In this work we formulate a continuum model of vitreous silica with its anomalous behavior based on infinitesimal strains. The model describe the frequently used sigmoidal curve of compression, which is seen here as an indication of a phase transformation between two phases at the macroscale. For simplicity and due to the aforementioned properties, we assume, that the phase transformation only affects the volumetric strain. Equally, we restrict the problem to an isothermal situation. \\
This work is structured as follows. Using an established variational formulation for inelastic materials, two energy potentials $\Psi$ (Helmholtz free energy) and $\Delta$ (Dissipation) are introduced first. 
After that, the evolution equations are calculated via a minimization principle, whereby a Lagrangian is minimized with respect to the internal or history variables.\\
We have to note, that the construction of the model is strongly related to our previous work published in \cite{DH}, see also \cite{DH26}. The present work is intended to elaborate an innovative application of the general formulation proposed there. Variational models for the evolution of inelastic materials, which are related to our approach, can be found in the literature, see for example \cite{OrtizStanierMethApplMechEng1999,MielkeContMechTD2003,MieheLambrecht2003,MieheLambrechtGuerses04,kochmann2011evolution} for applications to plasticity. Applications to phase transformations and shape memory alloys can be found in \cite{govindjee2007upper,waimann2016coupled}, and to pressure-dependent plasticity in \cite{Behr}. A novel variational model for microstructures in granular materials is presented in \cite{Khan2021}. Investigations of reciprocal and non-reciprocal thermodynamic extremal principles are found in \cite{HACKL2020103780} and \cite{HACKL2020104149}.

\section{Material model}
\label{sec:1}
\begin{figure}[h]
\centering
  \begin{minipage}[b]{.7\linewidth} 
   \centering
      \includegraphics[width=\linewidth]{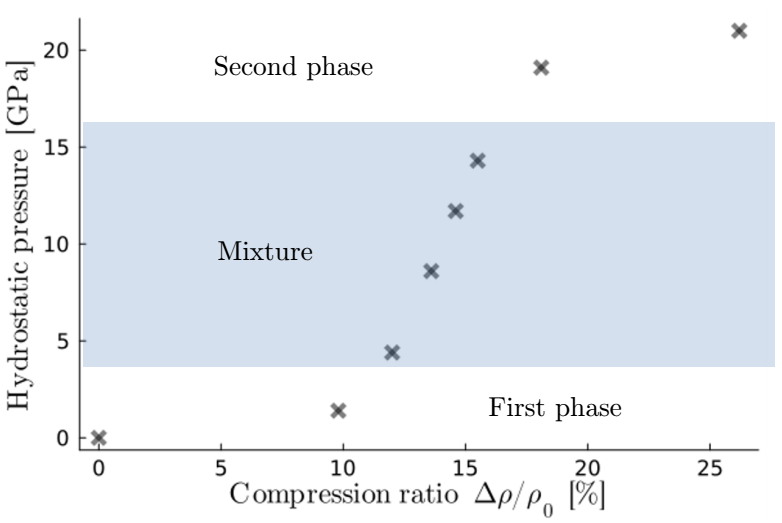}
      
   \end{minipage}
 \caption{Sigmoidal behavior of pressure a function of the compression ratio. The data set is taken from Deschamps et al. \cite{Deschamps14}. The suspected phase-mixing zone is highlighted in color. \raggedright }
      \label{fig:1}
\end{figure}

For the sake of simplicity, we assume an isotropic phase development of the material during the compression process. 
We further assume, that the transition between two phases  with phase fractions $\lambda_1 = 1-\lambda$ and $\lambda_2 = \lambda$ takes place within a fixed pressure range, here 0 GPa to 20 GPa. In between, two phases can coexist and possibly occur in microstructure zones such as shear bands. Figure \ref{fig:1} shows a sigmoidal behavior of the pressure a a function of the compression ratio. The experimental data, which are used as example are results from \textit{ex situ} measurements taken from Deschamps et al. \cite{Deschamps14}. We will show, that this behavior can be explained via the transition between two stable states of an energy potential, see Fig. \ref{fig:2}.\\

\subsection{ Helmholtz free energy}

The elastic energy part $\Psi_{\mathrm{elastic}}$ of the free energy potential can be written, considering infinitesimal strains, as 
\begin{align*}
\Psi_{\mathrm{elastic}}(\boldsymbol{\varepsilon},\boldsymbol{\varepsilon}^\mathrm{T}) = \frac{1}{2} (\boldsymbol{\varepsilon} -\boldsymbol{\varepsilon}^\mathrm{T} )\colon \mathbb{C} \colon (\boldsymbol{\varepsilon} -\boldsymbol{\varepsilon}^\mathrm{T})
\end{align*}
with the forth order stiffness tensor $\mathbb{C}$, the strain tensor $\boldsymbol{\varepsilon}$, and its decomposition into an elastic and transformation part
$\boldsymbol{\varepsilon} = \boldsymbol{\varepsilon}^\mathrm{e} + \boldsymbol{\varepsilon}^\mathrm{T}$.
To describe a material with several phases $k$, it should be noted that the associated strains $\boldsymbol{\varepsilon}$, $\boldsymbol{\varepsilon}^\mathrm{T}$ and the stiffness $\mathbb{C}$ may differ in each phase. In the general description we find,
\begin{align}
\label{eq02}
\Psi_{\mathrm{elastic}} = 	\sum_i^k \Psi^{\mathrm{elastic}}_i(\boldsymbol{\varepsilon}_i,\boldsymbol{\varepsilon}_i^\mathrm{T})
\end{align}	
with
\begin{align}
\Psi^{\mathrm{elastic}}_i(\boldsymbol{\varepsilon}_i,\boldsymbol{\varepsilon}_i^\mathrm{T}) = \frac{1}{2} (\boldsymbol{\varepsilon}_i -\boldsymbol{\varepsilon}_i^\mathrm{T} )\colon \mathbb{C}_i \colon (\boldsymbol{\varepsilon}_i -\boldsymbol{\varepsilon}_i^\mathrm{T})
\end{align}
and $i \in \{1, ... ,k\}$.\\
To analyse the energy of a phase mixture, we consider the average strain $ \boldsymbol{\varepsilon} = \sum_i \lambda_i \boldsymbol{\varepsilon}_i$ as constraint. Employing the Lagrange parameters and minimizing the elastic energy part with respect to $\boldsymbol{\varepsilon}_i$, we obtain the relaxed energy
\begin{align}
\Psi_{\mathrm{elastic}}^{\mathrm{rel}}(\boldsymbol{\varepsilon},\boldsymbol{\varepsilon}_{\mathrm{eff}}^\mathrm{T})= \min\{\Psi_{\mathrm{elastic}}~ \vert~ \boldsymbol{\varepsilon}_i ,~ \sum_i^k \lambda_i \boldsymbol{\varepsilon}_i~ = ~ \boldsymbol{\varepsilon}\}.
\end{align}
Minimization yields the macroscopic properties as follows:
\begin{align}
 \mathbb{C}_{\mathrm{eff}} &= \left(\sum_i^k \lambda_i \mathbb{C}_i^{-1}\right)^{-1}, ~~ \boldsymbol{\varepsilon}_{\mathrm{eff}}^\mathrm{T} = \sum_i^k \lambda_i \boldsymbol{\varepsilon}_i^\mathrm{T}.\label{eq:3}
\end{align}
Insertion into Eq.~\eqref{eq02} results in:
\begin{align}
	\label{eq06}
\Psi_{\mathrm{elastic}}^{\mathrm{rel}}(\boldsymbol{\varepsilon},\boldsymbol{\varepsilon}_{\mathrm{eff}}^\mathrm{T})= \frac{1}{2} (\boldsymbol{\varepsilon} -\boldsymbol{\varepsilon}_{\mathrm{eff}}^\mathrm{T} )\colon \mathbb{C}_{\mathrm{eff}}  \colon (\boldsymbol{\varepsilon} -\boldsymbol{\varepsilon}_{\mathrm{eff}}^\mathrm{T}) 
\end{align}
Reformulation of this energy term can be used to identify volumetric and deviatoric parts. 
In our case, considering that the only inelastic process is a volumetric phase transformation, Eq.~\eqref{eq06} is simplified to
\begin{align}
\Psi_{\mathrm{elastic}}^{\mathrm{rel}}(\boldsymbol{\varepsilon},\Theta,\lambda,\Theta^\mathrm{T}) = \frac{1}{2} \mathrm{K}_{\mathrm{eff}}(\Theta + \lambda \Theta^{\mathrm{T}})^2 + \mu_{\mathrm{eff}} \Vert \mathrm{dev} \boldsymbol{\varepsilon} \Vert ^2
\end{align}
where
\begin{align}
\mathrm{K}_{\mathrm{eff}} = \left(\sum_i^k \lambda_i \mathrm{K}_i^{-1}\right)^{-1}, \qquad \mu_{\mathrm{eff}} = \left(\sum_i^k \lambda_i \mu_i^{-1}\right)^{-1} \label{eq07}
\end{align}
are the effective bulk-modulus as well as the effective shear modulus.
Additionally, the volumetric strain is abbreviated as
$\Theta= \mathrm{tr}(\boldsymbol{\varepsilon})$. This follows from the relationship between the initial volume $ \mathrm{V_0}$ and the volume change $\mathrm{d}\mathrm{V} = \mathrm{V} - \mathrm{V_0}$,
\begin{align*}
\frac{\mathrm{d}\mathrm{ V}}{\mathrm{V_0}}= \mathrm{tr}(\boldsymbol{\varepsilon})= \Theta
\end{align*} 
see for example, \cite{Kermouche}. In addition, the volumetric transformation strain from the first phase is assumed to be  $\Theta^{\mathrm{T}}_1=0$ and from the second phase to be  $\Theta^{\mathrm{T}}_2=-\Theta^\mathrm{T}$ with $\Theta^\mathrm{T} > 0$. Therefore, $ -\lambda \Theta^{\mathrm {T}}$ can be seen as effective transformation strain $\Theta^\mathrm{T}_{\mathrm{eff}}$.
The maximum value of transformation strain is set equivalent to the compression ratio, which is given as $ \frac{\Delta \rho}{\rho_0}_{\mathrm{max}}$ = 21\%. This ratio describes the inelastic density change, as explained in \cite{Rouxel}.
If we consider a mixture of two different components, the macroscopic or actual elastic moduli can be calculated over the well-known \textit{rule of mixture} for binary materials, which agrees with Eq.~\eqref{eq:3} or Eq. \eqref{eq07}, respectively.  In our case the volume fractions $\lambda$ and $1 - \lambda$ as weighting proportions are used to find the effective property.\\ Additionally we assume that the stresses over each point in the amorphous structure remains constant. The minimum bulk modulus achieved in the classical \textit{Reuss lower bound} formulation is, 
\begin{align*}
\mathrm{K}_{\mathrm{Reuss}}=\frac{1}{2} \left(\frac{1-\lambda}{\mathrm{K}_{\mathrm{matrix}}}+\frac{\lambda}{\mathrm{K}_{\mathrm{vf}}}\right)^{-1}.
\end{align*} The moduli $\mathrm{K}_{\mathrm{matrix}}$ and $\mathrm{K}_{\mathrm{vf}}$ are stiffnesses for the matrix and for a different volume fraction, respectively. 
\begin{figure}[h]
\centering
  \begin{minipage}[b]{.6\linewidth} 
   \centering
      \includegraphics[width=\linewidth]{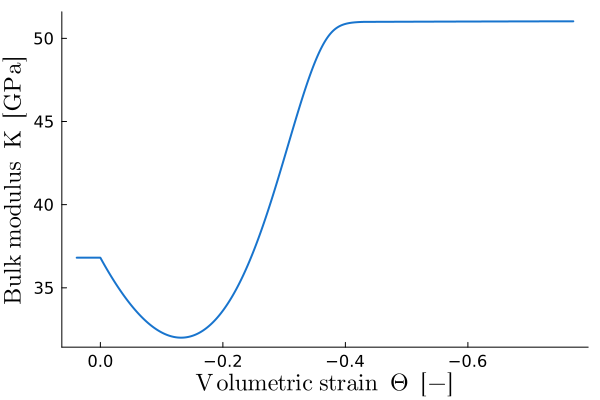}
      
   \end{minipage}
 \caption{Bulk modulus K against volumetric strain $\Theta$ with minimum between $0.0 < \Theta < 0.4 $, and saturation above $\Theta \geq 0.4$, see Eq. \eqref{eq:5}.  \raggedright }
      \label{fig:2}
\end{figure}
There are many possibilities to model the anomalous behavior of vitreous silica from a continuum mechanical view \cite{Li,Keryvin,Kermouche}.
We assume, that the physical effect of softening occurs due to the wide-mesh network structure and follow the idea of Birch et al., see \cite{Birch}. Then the elastic moduli can be specified depending on the relative density change $ \frac{\Delta \rho}{\rho_0}= - \lambda \cdot \Theta^\mathrm{T}$, or the strain $\Theta$. For example, in Deschamps et al. \cite{Deschamps14}, a quadratic function describes the behavior of the longitudinal modulus, and due to their the elastic moduli can also be represented by a similar expression. 
Therefore, the bulk modulus for the matrix can be replaced by K($\Theta$), which is a function of the volumetric strain. We suggest : \\
\begin{align}
\mathrm {K}(\Theta) = \left\{
\begin{array}{ll}
\mathrm{K}_1 & \Theta > 0 \\
\mathrm{K}_1 + (\mathrm{K}_1-\mathrm{K}_0) \frac{\Theta}{\Theta_\mathrm{a}} + \mathrm{K}_3 \Theta^2  & \, \textrm{otherwise}. \\
\end{array}
\right. \label{eq:5}
\end{align} 
In this case, the bulk modulus $\mathrm{K}_{\mathrm{matrix}}$ consist of three different parameters $\mathrm{K}_0$, $\mathrm{K}_1$ and $\mathrm{K}_3$. Whereas K($\Theta$) is set to a constant value in the tensile regime. To control the specific range of minimum value, an additional parameter $\Theta_\mathrm{a}$ $>$ 0 is introduced. Figure \ref{fig:2} shows the behavior of the bulk modulus $\mathrm{K}(\Theta)$, Eq. \eqref{eq:5}. The chosen values are listed in Tab. \ref{tab1}.

\subsubsection{Energy mixing term}
\begin{figure}[ht]
\centering
  \begin{minipage}[b]{.8\linewidth} 
   \centering
  
      \includegraphics[width=\linewidth]{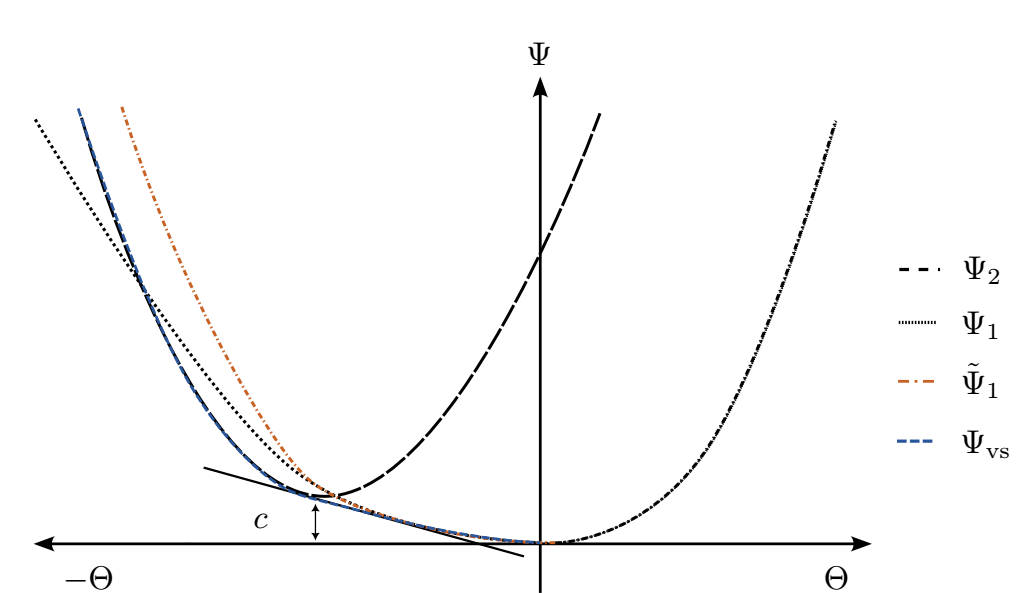}
       \end{minipage}
 \caption{Visualisation of potentials as function of volumetric strain $\Theta$. Unrelaxed potentials $\Psi_1, \Psi_2$ with energy shift c and tangent (solid line), $\Psi_{\mathrm{vs}}$ relaxed potential and $\tilde{\Psi}_1$, the corrected potential of $\Psi_1$. \raggedright }
      \label{fig:3}
\end{figure}

Based on the mixing entropy for ideal system of mixtures, we use a free energy mixing term $\Psi_{\mathrm{mix}}$ with amplitude $a$ $>$ 0. In the general setting for $k$ phases, this reads:
\begin{align*}
\Psi_{\mathrm{mix}}= a \sum_i^k \lambda_i \mathrm{ln}(\lambda_i).
\end{align*}
And for two phases that gives: 
\begin{align*}
\Psi_{\mathrm{mix}}= a((1-\lambda) \mathrm{ln} (1-\lambda) + \lambda \mathrm{ln}(\lambda)).
\end{align*}
Together with an energetic offset $c$, between both phases, also as chemical energy. The total energy is: 
\begin{align*}
\Psi_{\mathrm{tot}}= \Psi_{\mathrm{elastic}}^{\mathrm{rel}} + \Psi_{\mathrm{mix}} + \lambda c.
\end{align*}
Therefore, the Helmholtz free energy $\Psi_{\mathrm{vs}}$ for vitreous silica reads,
\begin{align}
\Psi_{\mathrm{vs}} = \frac{1}{2} \left(\frac{1-\lambda}{\mathrm{K}(\Theta)}+\frac{\lambda}{\mathrm{K}_2}\right)^{-1} (\Theta + \lambda \Theta^{\mathrm{T}})^2 + a((1-\lambda) \mathrm{ln} (1-\lambda) + \lambda \mathrm{ln}(\lambda))+ \lambda c.\label{eq:6}
\end{align}
Furthermore, the individual unrelaxed potentials can be written as:
\begin{align}
\tilde{\Psi}_1 = \hspace{0.1cm}& \frac{\mathrm{K(\Theta)}}{2} \Theta^2,\label{eq:7}\\ 
\Psi_2 = \hspace{0.1cm} & \frac{\mathrm{K}_2}{2}(\Theta + \Theta^\mathrm{T})^2.\label{eq:8}
\end{align}
For visualization see Fig.~\ref{fig:3}.
In the following, the subscript $^\mathrm{rel}$ for relaxed is omitted. 
\subsection{Dissipation potentials}

Starting from the general formulation introduced in  \cite{DH}, the internal dissipation can be described using the following approach:
\begin{align}
			\Delta_{\mathrm{reg}}&=  \frac{1}{2 \eta} \dot{\lambda}^2
			\\\label{eq:14}
			\Delta_{\mathrm{trans}}&= \dot{ \lambda}~D( \Theta^\mathrm{T} ).
\end{align}
In this case, $D$ is the called \textit{dissipation distance} \cite{Mielke}, which describes an immediate transition. The rate $\dot{ \lambda}$ influences the process of phase transition between the two phases. 
Following this, the \textit{dissipation distance}
reads, $D(  \Theta^\mathrm{T})=r \vert\Theta^\mathrm{T} \vert$, with $r$ a dissipation constant, $\eta$ viscosity value. \\
\subsection{Evolution equation}
To gain access to microstructure development the principle of minimum of the dissipation potential then leads to the equation
\begin{align*}
\min\limits_{ \{\dot{\lambda}\ge 0\}} \left(  \dot \Psi_{\mathrm{vs}} + \Delta_{\mathrm{reg}} + \Delta_{\mathrm{trans}} \right).
\end{align*}
The corresponding Lagrangian is the summation of the total derivative of $\Psi_{\mathrm{vs}}$ with respect to time and the dissipation potentials 
\begin{align*}
\mathcal{L}=  -q \cdot \dot \lambda  
 + \dot{\lambda} r \vert  \Theta^\mathrm{T} \vert  + \frac{1}{2\eta} \dot{\lambda}^2.
\end{align*}
Minimisation with respect to the internal variable $\dot{\lambda}$, leads to the stationarity condition:
\begin{align}
0 \in \frac{\partial \mathcal{L}}{\partial \dot{\lambda}} &\hspace{1cm} \Rightarrow \hspace{1cm}\dot{\lambda} =  \eta (q  + r \vert  \Theta^\mathrm{T} \vert  )_{+} \label{eq:15}
\end{align}
This resulting differential equation for $\dot{\lambda}$ Eq. \eqref{eq:15}, is explicit and involves a yield condition $f = q + r \vert  \Theta^\mathrm{T} \vert$ for phase development, where $\mathrm{max}(b,0)=(b)_+$.\\
A thermodynamical driving force $q$ for phase change can be calculated via the equation
\begin{align*}
q &=- \frac{\partial \Psi_{\mathrm{vs}}}{\partial \lambda} \\ &= - \biggl[a(- \mathrm{ln} (1-\lambda) +  \mathrm{ln}(\lambda))-\frac{1}{2}\left(-\frac{1}{\mathrm{K}(\Theta)}+ \frac{1}{\mathrm{K}_2} \right) \left(\frac{1-\lambda}{\mathrm{K}(\Theta)}+ \frac{\lambda}{\mathrm{K}_2} \right)^{-2} \\&(\Theta +\lambda \Theta^\mathrm{T} )^2 
+\left(\frac{1-\lambda}{\mathrm{K}(\Theta)} + \frac{\lambda}{\mathrm{K}_2} \right)^{-1} \Theta^\mathrm{T} ( \Theta + \lambda \Theta^\mathrm{T}) + c \biggl].
\end{align*}
Besides this, the first and second derivatives from the Helmholtz free energy potential $\Psi$ 
\begin{align}
\mathrm{P} = - \frac{\mathrm {d} \Psi}{ \mathrm{d} \mathrm{V}/\mathrm{V_0}} \hspace{0.2cm}, \hspace{0.2cm} \mathrm{K} = -  \frac{\mathrm{d} \mathrm{P}}{ \mathrm{d} \mathrm{V}/\mathrm{V_0}}  
\end{align}
are then used to find the related hydrostatic pressure $\mathrm{P}$ and the bulk modulus $\mathrm{K}$.
Now we are able to identify the hydrostatic pressure $\mathrm{P}$ as written before
\begin{align*}
\mathrm{P} &= - \frac{\partial \Psi_{\mathrm{vs}}}{\partial \Theta} \\&= -\biggl(\frac{1}{2}(1-\lambda) \frac{\partial \mathrm{K}(\Theta)}{\partial \Theta} \mathrm{K}(\Theta)^{-2} \left(\frac{1-\lambda}{\mathrm{K}(\Theta)} + \frac{\lambda}{\mathrm{K}_2} \right)^{-2}(\Theta +\Theta^{\mathrm{T}}_{\mathrm{eff}} )^2 + (\Theta + \Theta^{\mathrm{T}}_{\mathrm{eff}})\\& \left(\frac{1-\lambda}{\mathrm{K}(\Theta)} + \frac{\lambda}{\mathrm{K}_2} \right)^{-1}\biggl).
\end{align*}
A relationship for the hydrostatic pressure $\mathrm{P}$ and the hydrostatic stress $\sigma_{\mathrm{h}}$, is $\mathrm{P}= -\sigma_{\mathrm{h}}$. 
Furthermore to get a better understanding of the 2d behavior of our model the general formula of Hooke's law 
\begin{align}
\boldsymbol{\sigma} = - \mathrm{P} \boldsymbol{\mathrm{I}} + 2\mu_\mathrm{eff}~ \mathrm{dev} (\boldsymbol{\varepsilon})\label{eq:17}
\end{align} can be used for numerical consideration. The parameter $\mathrm{\mu_{eff}}$ is the effective shear modulus. And the effective bulk modulus $\mathrm{K}_{\mathrm{eff}}$ depends only on the volumetric strain and the evolution of the developing phase
\begin{align*}
\mathrm{K}_{\mathrm{eff}}= - \frac{\partial \mathrm{P}}{\partial \Theta} = \left(\frac{1-\lambda}{\mathrm{K}(\Theta)} + \frac{\lambda}{\mathrm{K}_2} \right)^{-1}.
\end{align*}

\section{Numerical experiments}

First, we start with simulations of the material point behavior in Subsection \ref{subs1}. After that, we will focus on the energy potentials and the connection of the most common diagrams getting out of pseudo - hydrostatic experiments using a Diamond Anvil Cell (DAC). Then, 2d Finite Element tests will demonstrate the numerical behavior of the model under different loading conditions; in Subsection \ref{subs2}. The simulations include an indentation test, in \ref{subsubs1}, an uniaxial compression test \ref{subsub2} and a coupled shear-compression test, in \ref{subsubs3}. Finally, as a benchmark problem, a square plate with a circular hole is featured, in \ref{subsubs4}. \
The numerical simulations presented here are done with the open source programming language \emph{Julia}. In addition, the package \emph{Julia Ferrite} in combination with \emph{Gmsh} is used for the FE calculations. The created unstructured meshes consist of quadrilateral elements with four quadrature points. For the following examples, material parameters were chosen that are approximately based on the experimental data from Deschamps et al. (2014)  \cite{Deschamps14}.

\subsection{Material point behavior }\label{subs1}

\begin{figure}[h]
\centering
  \begin{minipage}[b]{.45\linewidth} 
   \centering
      \includegraphics[width=\linewidth]{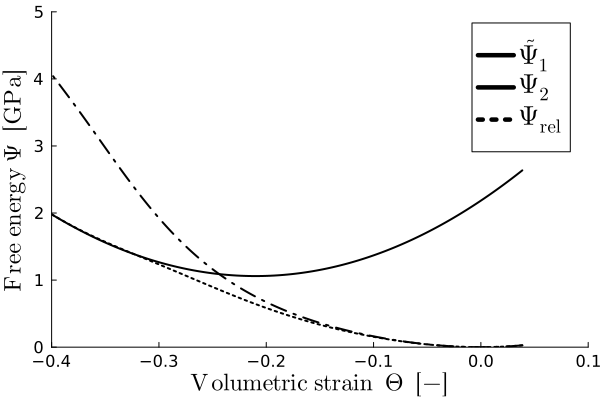}
      (a)
   \end{minipage}
    \begin{minipage}[b]{.45\linewidth} 
   \centering
      \includegraphics[width=\linewidth]{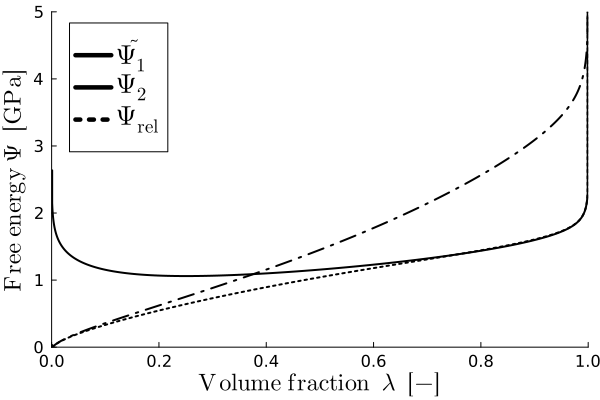}
       (b)
   \end{minipage}
   
 \caption{Development of separated energy potentials $\tilde{\Psi}_{1},\Psi_{2}$  from Eq. \eqref{eq:7} and Eq. \eqref{eq:8}, with evolution of the relaxed energy potential $\Psi_{\mathrm{vs}}$, Eq. \eqref{eq:6}.  \raggedright }
      \label{fig:4}
\end{figure}

\begin{figure}[h]
\centering
   \begin{minipage}[b]{.45\linewidth} 
   \centering
      \includegraphics[width=\linewidth]{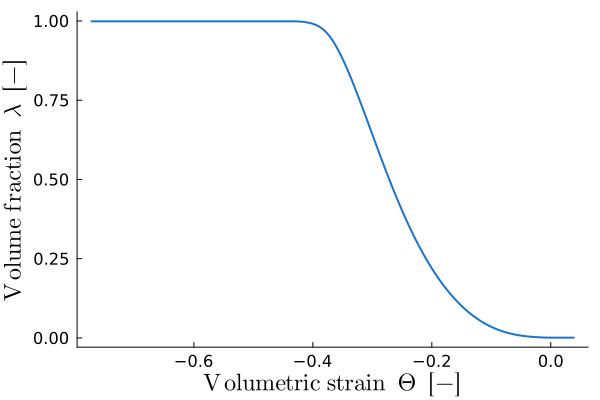}
      (a) 
   \end{minipage}
  \begin{minipage}[b]{.5\linewidth} 
   \centering
      \includegraphics[width=\linewidth]{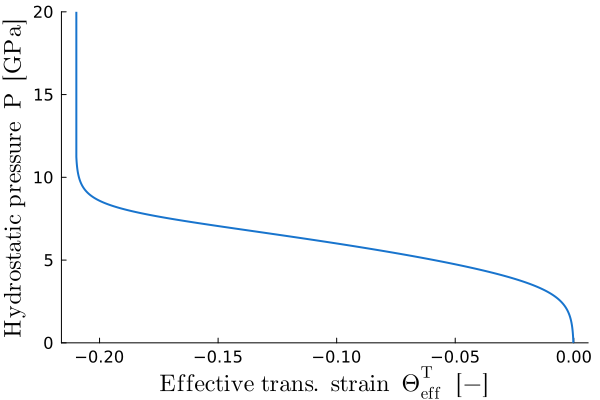}
      (b)
   \end{minipage}
   \begin{minipage}[b]{.5\linewidth} 

    \centering
      \includegraphics[width=\linewidth]{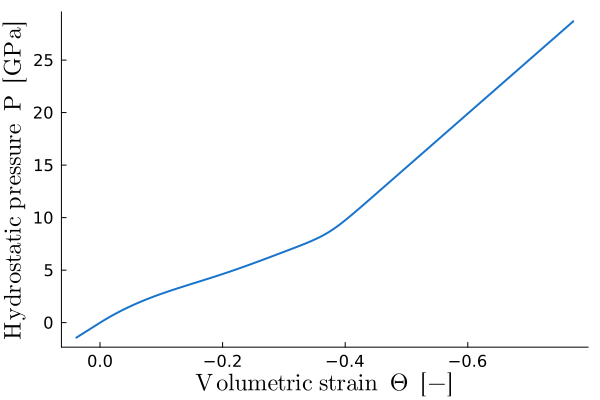}
      (c)
   \end{minipage}
\centering
  \begin{minipage}[b]{.45\linewidth} 
   \centering
      \includegraphics[width=\linewidth]{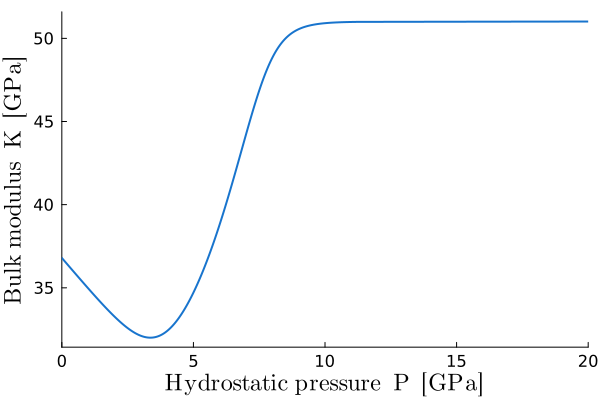}
      (d)
   \end{minipage}

   \caption{Isothermal material point analysis. From (a) to (d), volume fraction as function of volumetric strain, EOS - function, pressure-compression correlation, bulk modulus against pressure. \raggedright }
      \label{fig:5}
\end{figure}

The individual energy potentials, Eq.~\eqref{eq:7} and Eq.~\eqref{eq:8}, are shown in Fig.~\ref{fig:4} together with the relaxed energy, see Eq.~\eqref{eq:6}. In order to include the experimentally achieved densification ratio of $ \frac{\Delta \rho}{\rho_0}$ = 21\%, the volumetric transformation strain is set to be $\Theta^\mathrm{T}$ = 0.21. Further material parameters are listed in Tab.~\ref{tab1} and \ref{tab1.1}. For the following material point analyses the volumetric change during compression -$\Theta$ varies in a range of (-0.1, 0.4) employing 200 load steps.
In this numerical example the volume fractions are allowed to increase from $\lambda = 0.001$ to $\lambda= 0.999$. Figure \ref{fig:5} (a) shows a phase transformation between a volumetric change from 0 to -0.4. The EOS - function, hydrostatic pressure $\mathrm{P}$ as function of effective transformation volume change $\Theta^{\mathrm{T}}_{\mathrm{eff}}$, Fig \ref{fig:5} (b), indicates a transition between 5 GPa and 12 GPa. The pressure values are lowered by (5-7) GPa compared to \cite{Deschamps14} or other DAC-experiments and is a result by the parameters used. Phase development starts around 5 GPa, and the transformation is faster compared to experimental results. In addition it has a different curvy shape at the interval $\Theta^\mathrm{T}_{\mathrm{eff}}$ from 0.00 [-] to -0.05 [-] and from -0.15 [-] to -0.21 [-].

\begin{table}[h]
\centering
\caption{Bulk modulus values for material point simulation.  }\label{tab1}%
\begin{tabular}{@{}   c | c | c | c  @{}}
\hline

$\mathrm{K_0}$ [GPa] & $\mathrm{K_1}$ [GPa]& $\mathrm{K_2}$ [GPa]& $\mathrm{K_3}$ [GPa]\\
\hline

 32.0 &36.8  & 51.0& 185.1 \\

\hline
\end{tabular}
\end{table}

\begin{table}[h]
\centering
\caption{Parameters for material point simulation. }\label{tab1.1}%
\begin{tabular}{@{}   c | c | c | c | c  | c   @{}}
\hline

$\mathrm{r}$ [GPa]&$a$ [GPa] & $\Theta_{\mathrm{T}}$ [-]& $\Theta_{\mathrm{a}}$ [-] & $\eta$ [Pa]&$c$ [GPa]\\
\hline

4$\cdot 10^{-4}$& 0.14 & 0.21  & 7.1 $\cdot 10 ^{-2}$& 0.7 $\cdot 10^{11}$&1.6 \\

\hline
\end{tabular}
\end{table}
This behavior also can be seen in Fig. \ref{fig:5} (c) and (d). The hydrostatic pressure as function of volumetric strain Fig. \ref{fig:5} (c) shows a monotonic increase as also described in \cite{Schill1}. Whereas the volumetric behavior becomes linear elastic at higher pressure in contrast to MD-simulations \cite{Schill1}, without a stronger limitation of volume change. The anomalous softening effect of the bulk modulus is presented in Fig. \ref{fig:5} (d). The minimum is reached at around 4 GPa, which is a higher value compared to experimental data and a lower value as has been found by MD-Simulations \cite{Mysen, Deschamps14, Huang1}. The transformation terminates at approx. 13 GPa.\\
\begin{figure}[h]
\centering
  \begin{minipage}[b]{.6\linewidth} 
 
      \includegraphics[width=\linewidth]{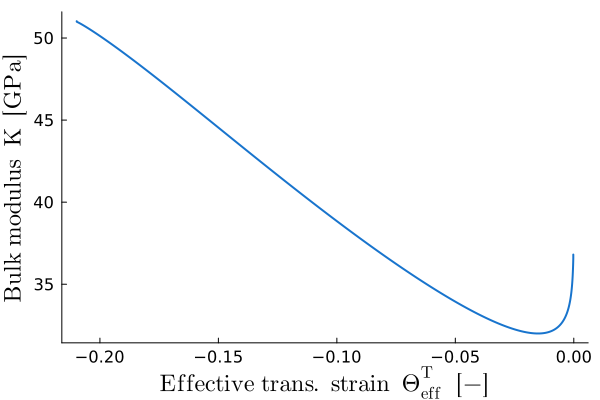}
      
   \end{minipage}
 
 \caption{Bulk modulus K as function of effective transformation strain $\Theta^{\mathrm{T}}_{\mathrm{eff}}$.  \raggedright }
      \label{fig:6}
\end{figure} 
The assumption of an isotropic phase transformation in an ideal body at  macroscopic level, leads in general to the formula,
\begin{align}
\mathrm{c}_\mathrm{l}^2 = \frac{\mathrm{K}+ \frac{3}{4}\mu }{\rho},\label{eq:17.1}
\end{align} for the longitudinal sound velocity, see  \cite{Wang,Deschamps14,Merklein,Grimsditch}. In addition to other ways of measuring the speed of sound in solids, Brillouin scattering is often used as a non-destructive method, which allows access to mechanical properties of glasses during high pressure studies. It describes in general inelastic interactions of light with an acoustic wave in a material, whereby this acoustic wave can be longitudinal (fluctuations in density) or transverse.\\
A strong connection of the longitudinal sound velocity and the bulk modulus, is known from the kinetic theories of viscous flow in liquids \cite{Wang}, using the formula $\mathrm{c}_\mathrm{l}^2 = \frac{\mathrm{K} }{\rho}$. Therefore, Fig.~\ref{fig:6} presents the bulk modulus as function of effective transformation strain, as it is documented in \cite{Deschamps14}.
Starting at around 36 GPa, Fig.~\ref{fig:6} demonstrates the softening effect. Although this behavior is in contrast to data gathered from \textit{ex situ} experiments in \cite{Deschamps14}, it illustrates the distinctive relationship between the behavior of all curves shown in Fig.~\ref{fig:5}. 
Otherwise, Eq.~\eqref{eq:17} allows access to Poisson's ratio or the deviatoric stresses, for example, if the strain is known, and the longitudinal sound velocity to be determined according to Eq.~\eqref{eq:17.1}.

\subsection{Finite element results }\label{subs2}

\begin{figure}[h]
\centering
  \begin{minipage}[b]{.9\linewidth} 
   \centering
       \includegraphics[width=\linewidth]{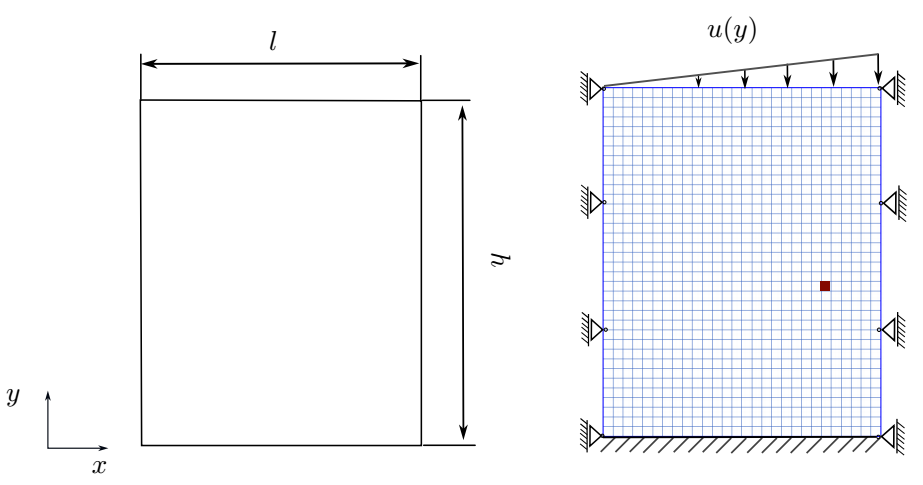}
   \end{minipage}
 \caption{Illustration of the two-dimensional geometry on the x - y plane on the left side and the used mesh and boundary values on the right. One element selected for comparison of material point behavior is colored dark red. \raggedright }
      \label{fig:7}
\end{figure}
\subsubsection{ Compression test}\label{subsubs1}
Figure \ref{fig:7} illustrates the geometric setting with loading conditions of a compression test. A rectangular plate with its length  $l = 2$ \ce{mm} and height $h = 2.5$ \ce{mm} is subject to a triangular displacement $u(y)$ at the upper boundary. In addition, the displacement $u_x$ is fixed on the left and right boundary, while the plate is clamped at the bottom. I.e., the boundary values are given as :
\begin{align*}
u_x &= 0~\mathrm{for}~x = 0  \\
u_x &= 0~\mathrm{for}~x = l \\
 u_y &= u_x = 0~\mathrm{for}~y = 0 .
\end{align*}
A quadrilateral mesh with 1136 elements is used for the simulation. The chosen material parameters are listed in Tab.~\ref{tab2} and \ref{tab2.2}.
\begin{figure}[h]
\centering
  \begin{minipage}[b]{.85\linewidth} 
   \centering
       \includegraphics[width=\linewidth]{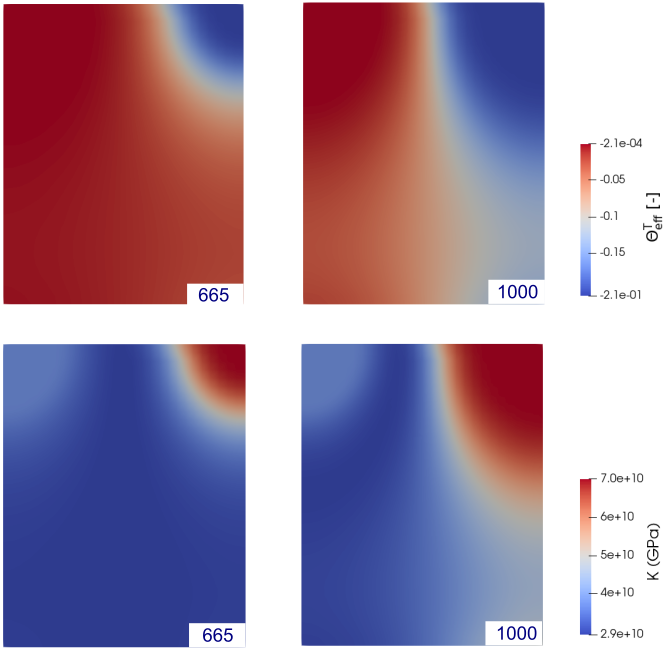}
   \end{minipage}
 \caption{Representation of spreading bulb at two time steps n = [665, 1000]. Comparison of bulk modulus K and effective transformation strain $\Theta^{\mathrm{T}}_{\mathrm{eff}}$. \raggedright }
      \label{fig:8}
\end{figure}
\begin{table}[h]
\centering
\caption{ Moduli for the 2d compression test. }\label{tab2}%
\begin{tabular}{@{} c | c | c | c  | c @{}}
\hline

$\mathrm{K_0}$ [GPa] & $\mathrm{K_1}$ [GPa]& $\mathrm{K_2}$ [GPa]& $\mathrm{K_3}$ [GPa]& $\mu$ [GPa]\\
\hline

 30.0 &36.0  & 70.0& 170.0&30.0 \\

\hline
\end{tabular}
\end{table}
\begin{table}[h]
\centering
\caption{ Material parameters for the 2d compression test. }\label{tab2.2}%
\begin{tabular}{@{} c | c | c | c | c  | c   @{}}
\hline

$r_1$ [GPa]&$a$ [GPa] & $\Theta_{\mathrm{T}}$ [-]& $\Theta_{\mathrm{a}}$ [-] & $\eta$ [Pa]&$c$ [GPa]\\
\hline

2$\cdot 10^{-4}$& 8.0 & 0.21  & 0.086 & 3 $\cdot 10^{9}$&1.0 \\

\hline
\end{tabular}
\end{table}

Figure \ref{fig:8} demonstrates a contour plot of the spreading bulb at two steps n = [665, 1000] from a test with a maximum of  $\mathrm{n}_{\mathrm{max}}$ = 1600 steps, while the animation time is t = 2$\si{\micro\second}$. The applied load is -0.99 $\si{\mega\pascal}$. The phase transformation begins in the upper right corner and a minimum of the bulk modulus appears to precede the densification process. Furthermore Fig. \ref{fig:8}, shows at the upper left hand side, the evolution of $\Theta^{\mathrm{T}}_{\mathrm{eff}}$, with the blue colored irreversible densified zone. A similar development trend of densification can be observed in indentation tests, see for example Bruns et al.\cite{Bruns}.
\begin{figure}[h]
\centering
  \begin{minipage}[b]{.6\linewidth} 
   \centering
       \includegraphics[width=\linewidth]{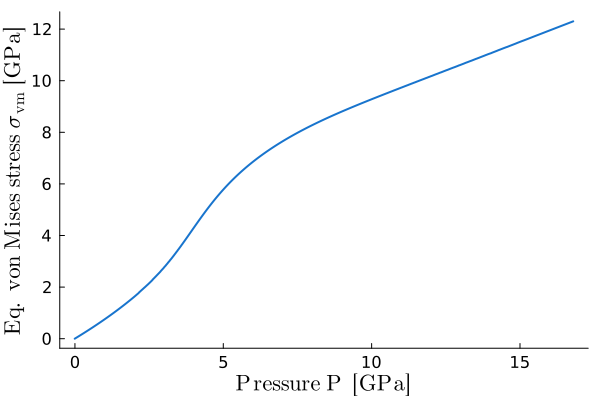}
   \end{minipage}
 \caption{Equivalent von Mises stress against hydrostatic pressure P. \raggedright }
      \label{fig:9}
\end{figure}
\begin{figure}[tp]
\centering
   \begin{minipage}[b]{.5\linewidth} 
   \centering
      \includegraphics[width=\linewidth]{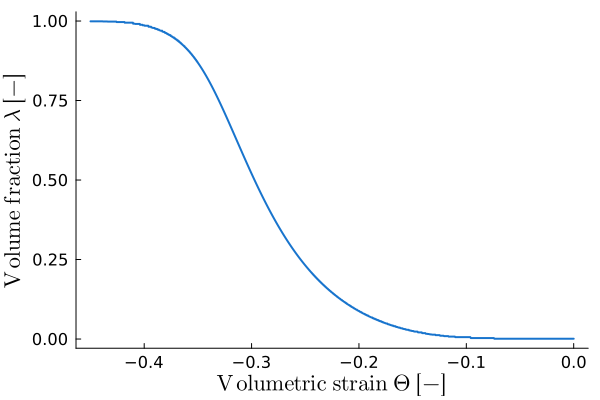}
      (a) 
   \end{minipage}
  \begin{minipage}[b]{.45\linewidth} 
   \centering
      \includegraphics[width=\linewidth]{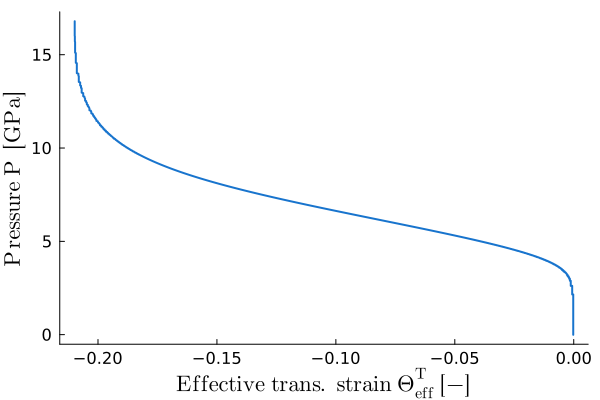}
      (b)
   \end{minipage}
   \begin{minipage}[b]{.45\linewidth} 
   \centering
      \includegraphics[width=\linewidth]{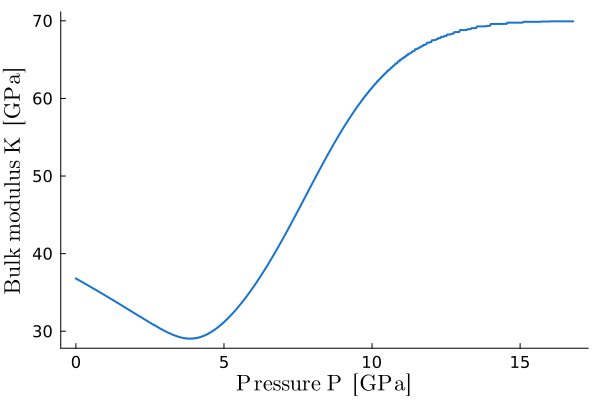}
      (c) 
   \end{minipage}
  \begin{minipage}[b]{.45\linewidth} 
   \centering
      \includegraphics[width=\linewidth]{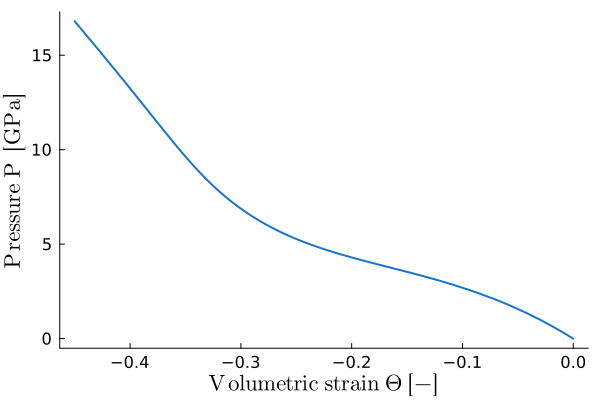}
      (d)
   \end{minipage}
   
\begin{minipage}[b]{.45\linewidth} 
   \centering
      \includegraphics[width=\linewidth]{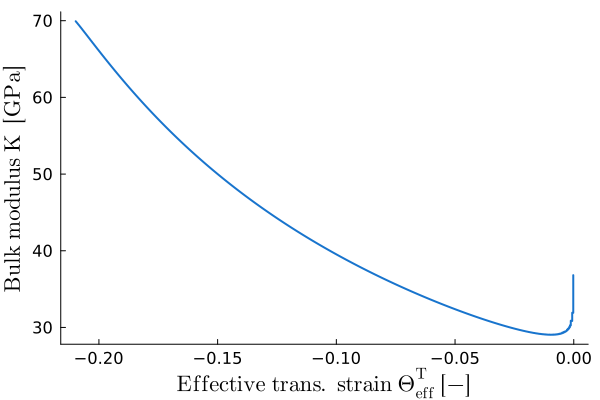}
       (e)
   \end{minipage}
\begin{minipage}[b]{.45\linewidth} 
   \centering
      \includegraphics[width=\linewidth]{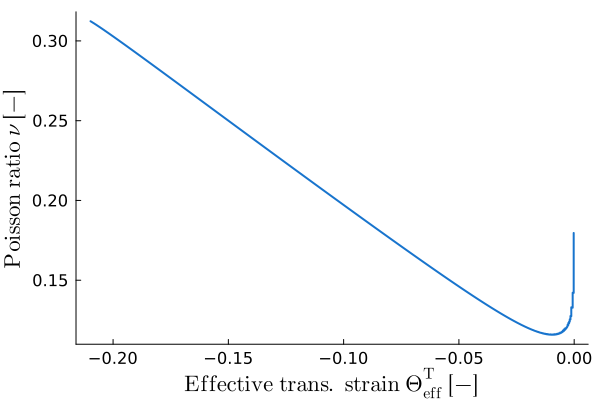}
       (f)
   \end{minipage}
   \caption{Progression of the model at one Gaussian point. Data are taken from the 2d FE compression test from one cell, see Fig. \ref{fig:7}. (a) Shows the developing phase $\lambda$ against the volumetric strain $\Theta$. From (b) to (f), equation of state (EOS), bulk modulus $\mathrm{K}$ against hydrostatic pressure $\mathrm{P}$, volumetric strain $\Theta$ as a function of hydrostatic pressure $\mathrm{P}$, bulk modulus $\mathrm{K}$ as a function of effective transformation strain $\Theta^{\mathrm{T}}_{\mathrm{eff}}$ and (f) Poisson ratio $\nu$ against effective transformation strain $\Theta^{\mathrm{T}}_{\mathrm{eff}}$. \raggedright }
      \label{fig:10}
\end{figure}
For completeness, results in Fig.~\ref{fig:9} and \ref{fig:10} show the behavior in the two dimensional FE compression test at one Gaussian point. An element in the middle is randomly selected for data extraction. It is shaded dark in Fig.~\ref{fig:7}. 
In the context of indentation tests, Fig. \ref{fig:9} displays the calculated von Mises stress, $\sigma_{\mathrm{vm}} = \sqrt{\frac{2}{3} s_{ij} s_{ij}}$ plotted against hydrostatic pressure P. The behavior and range of values show differences from other representations starting at a pressure of approx. P $>$ 10 $\si{\giga\pascal}$ \citep{Bruns}. Here a linear relationship is observed instead of a stress limitation, which is approx. at 8 GPa \cite{Molnar} for silicate glasses. This correlates with Fig.~\ref{fig:10} (d), which demonstrates the aforementioned monotonic increase of the pressure in relationship to the evolution of volume change. This could already indicate additional effects on an atomic scale at this level. However, the progress of the von Mises stress is comparable with results shown in \cite{Bruns}.\\
The phase transformation associated with compaction can be seen in Fig.~\ref{fig:10} (a). It starts at approx. $\Theta = -0.1$ and ends at approx. $\Theta = -0.4$. For further illustration, Figure  \ref{fig:10} (b)-(d) presents the well-known curves of silica glass with its permanent densification. The curves are largely identical to those from the material point analysis, see Section \ref{subs1}.
In addition Fig.~\ref{fig:10} (f) shows the associated poisson ratio $\nu$ against hydrostatic pressure P, calculated over the relationship between the moduli, K and $\mu$
 \begin{align}
 \nu = \frac{3 \mathrm{K} - 2 \mu}{6 \mathrm{K} + 2 \mu}.
\end{align}

\subsubsection{ Uniaxial compression test} \label{subsub2}
For completeness, our results at material point level are presented together with some experimental data in our last example.
\begin{figure}[h]
\centering
  \begin{minipage}[b]{.9\linewidth} 
   \centering
       \includegraphics[width=\linewidth]{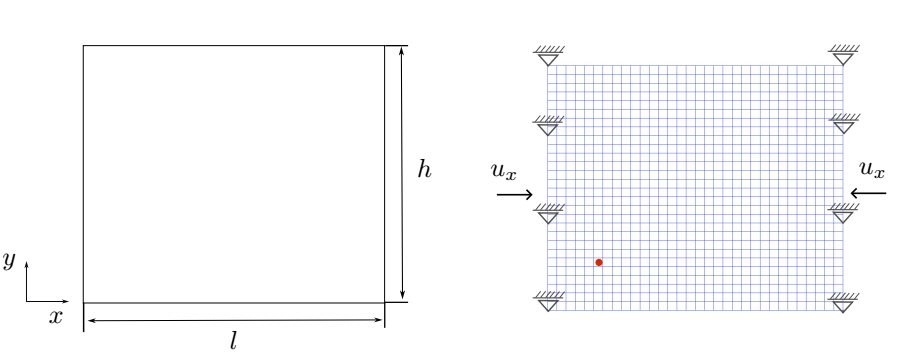}
   \end{minipage}
 \caption{Geometric setting for the uniaxial compression and the coupled shear - compression test, see Fig. \ref{fig:15}. On the left, the two-dimensional rectangle in the x - y plane, with length l = 1 \si{\micro\metre} and height h = 0.85 \si{\micro\metre}. On the right, mesh with 1016 elements and boundary conditions for the uniaxial compression test. The red marked element is used for data extraction, see Fig. \ref{fig:12}. \raggedright }
      \label{fig:11}
\end{figure}
Experimental results for silica glass indicates a development of $\mu$ from 30 to 45 \si{\giga\pascal} during the desification process \cite{Deschamps14, Schill1}, similar to the behavior of the bulk modulus K.
The following figures illustrate both situations. Simulations were performed with a constant shear modulus $\mu$ = 30 \si{\giga\pascal}, as well as with a shear modulus that varies as a function of $\Theta$, given by $\mu(\Theta) = 30.0 + b\cdot\Theta + a\cdot \Theta^2$, to allow a slight variation.\\

\begin{figure}[tp]
\centering
   \begin{minipage}[b]{.45\linewidth} 
   \centering
      \includegraphics[width=\linewidth]{
      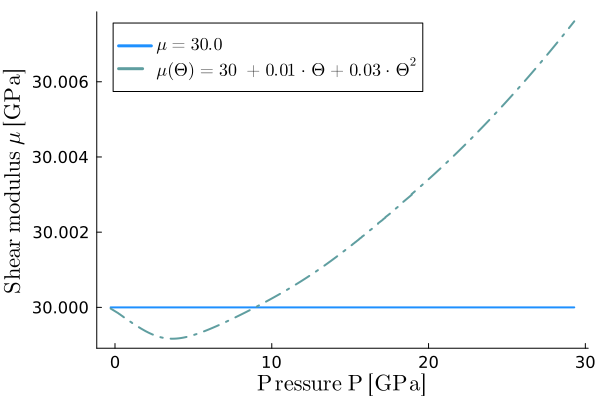}
      (a) 
   \end{minipage}
  \begin{minipage}[b]{.45\linewidth} 
   \centering
      \includegraphics[width=\linewidth]{
      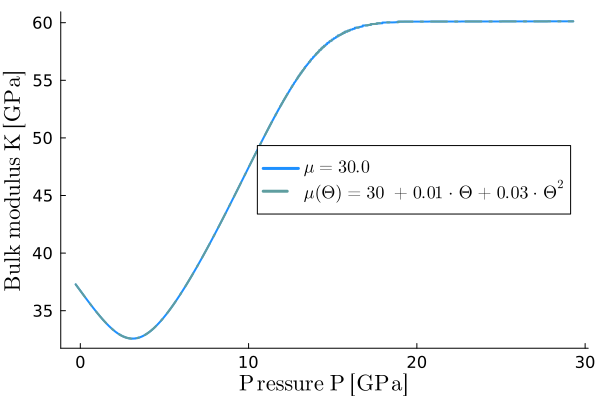}
      (b)
   \end{minipage}
   \begin{minipage}[b]{.45\linewidth} 
   \centering
      \includegraphics[width=\linewidth]{
      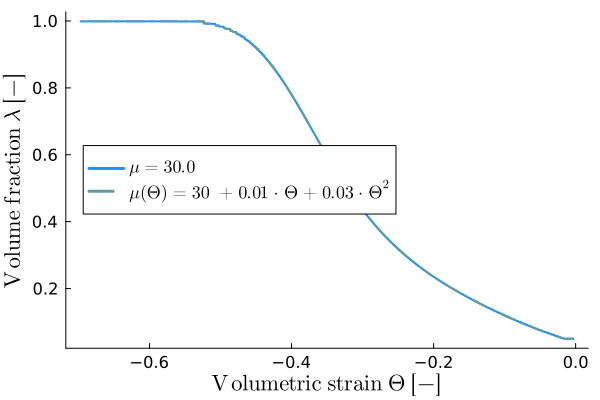}
      (c) 
   \end{minipage}
 \begin{minipage}[b]{.45\linewidth} 
   \centering
      \includegraphics[width=\linewidth]{
     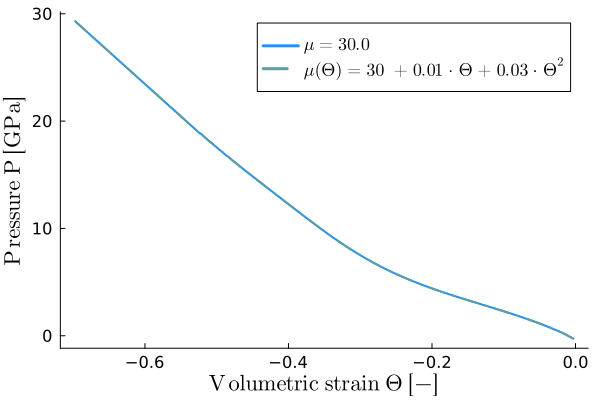 }
      (d) 
   \end{minipage}
   \begin{minipage}[b]{.45\linewidth} 
   \centering
      \includegraphics[width=\linewidth]{
      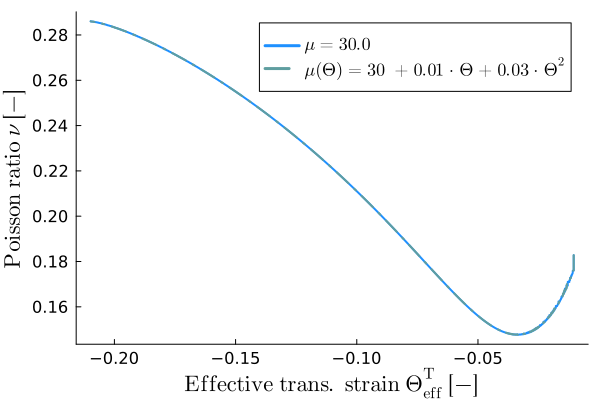}
      (e) 
   \end{minipage}
   \begin{minipage}[b]{.45\linewidth} 
   \centering
      \includegraphics[width=\linewidth]{
     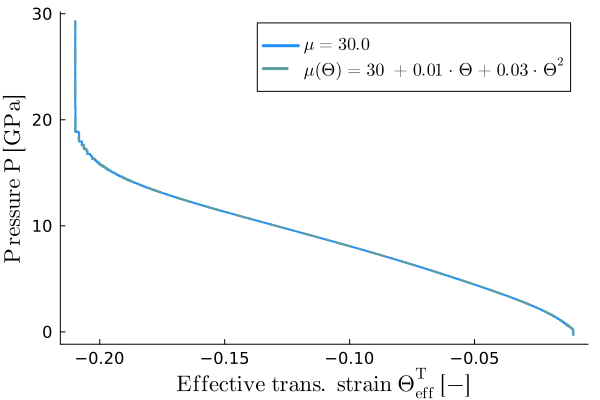}
      (f) 
   \end{minipage}
   \caption{Behavior of a single Gaussian point during compression. Comparison of two assumed shear moduli. (a) shear modulus $\mu (\Theta)$ as function of $\Theta$, (b) bulk modulus K, (c) increasing volume fraction $\lambda$, (d) pressure curve as function of  volumetric strain $\Theta$ (e)  poisson ratio $\nu$, (f) hydrostatic pressure P against effective transformation strain $\Theta^{\mathrm{T}}_{\mathrm{eff}}$.   \raggedright }
      \label{fig:14}
\end{figure}

To simulate an uniaxial compression test, the geometric setting and the boundary conditions in Fig.~\ref{fig:11} are applied.
\begin{align*}
u_y &= 0~\mathrm{for}~ x = 0~\mathrm{and}~ x = l \hspace{0.1cm} \\
u_x &=  \tilde{u}_x(t)~\mathrm{for}~ x = 0\\
u_x &= - \tilde{u}_x(t)~\mathrm{for}~ x = l .
\end{align*}
The length of the displayed sample is l = 1 \si{\micro\metre} and the height is h = 0.85 \si{\micro\metre}. The displacement is fixed in y - direction on the left side for $x = 0$ and and for $x = l$, whereas the material is subjected to a compressive displacement in x - direction at both boundaries. 
\begin{figure}[h]
\centering
  \begin{minipage}[b]{.9\linewidth} 
   \centering
       \includegraphics[width=\linewidth]{
     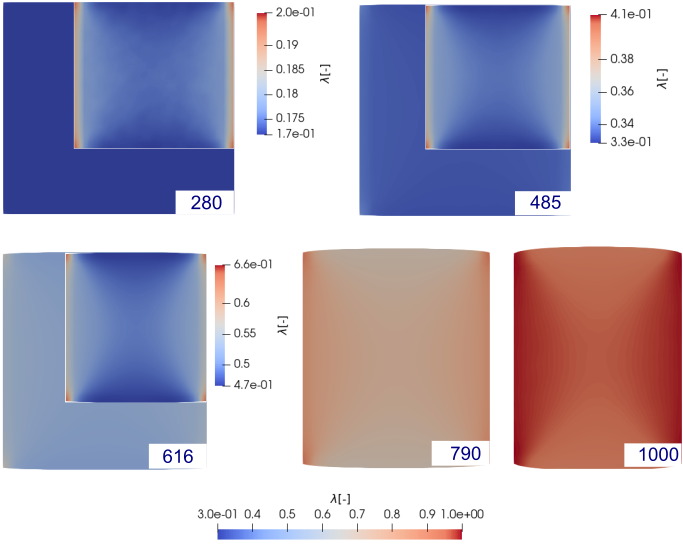}
   \end{minipage}
 \caption{Contour plot of phase transformation during uniaxial compression at five time steps, n = [280, 485, 616, 790, 1000]. The newly formed volume fraction starts to develop from $\lambda = 0.05$ to $\lambda = 0.999 $ at the corners. Three plots for the time step n =[280, 485, 616] show insets with refined color bars for illustrative purposes. To demonstrate the deformation, a factor of k = 0.6, is applied. \raggedright }
      \label{fig:12}
\end{figure}
An quadrilateral mesh with 1016 elements is used.
Material parameters are listed in Tab.~\ref{tab5}, \ref{tab5.1}. Figure \ref{fig:12} shows a contour plot of volume fraction development at five number of time steps n = [280, 485, 616, 790, 1000]. Over time, the newly created phase increasing within the range $0.05 \leq \lambda \leq 0.999$. Three insets at n = [280, 485, 616], shows the underlying development of the newly created phase. \\
In addition, the corresponding and consistent development of bulk modulus K, hydrostatic pressure P, volumetric strain $\Theta$ and phase $\lambda$ at four time steps n is displayed in Fig.~\ref{fig:13}. 
For this example a constant shear modulus $\mu = 30.0$ GPa is used.
In order to allow the shear modulus $\mu$ to have a slight dependence on the total strain $\Theta$ based on Eq. \ref{eq:5}, Fig.~\ref{fig:14} displays the comparison of both cases, at one Gaussian point from an element is marked with a red point in Fig.~\ref{fig:11}. We consider that this slight variance of shear modulus does not effect the behavior of the numerical solutions. Therefore in the following examples only the behavior with constant shear modulus is shown. Despite the small influence, we observe a minimum value of the shear modulus between $ 2 \leq \mathrm{P} \leq 5 $ [GPa] as expected, even though the parameters a = 0.03 and b = 0.01 for $\mu(\Theta)$ were chosen arbitrarily.  
\begin{figure}[h]
\centering
   \begin{minipage}[b]{.45\linewidth} 
   \centering
      \includegraphics[width=\linewidth]{
      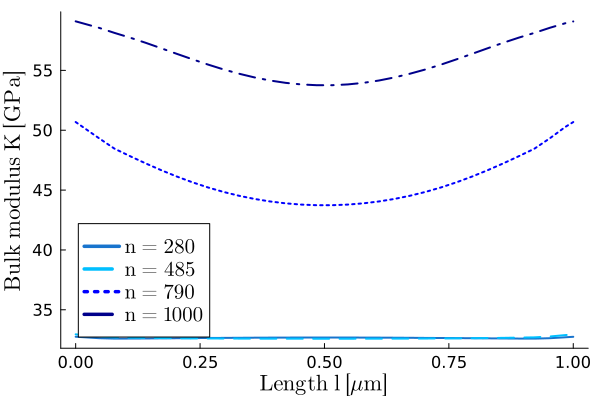}
      (a) 
   \end{minipage}
  \begin{minipage}[b]{.45\linewidth} 
   \centering
      \includegraphics[width=\linewidth]{
      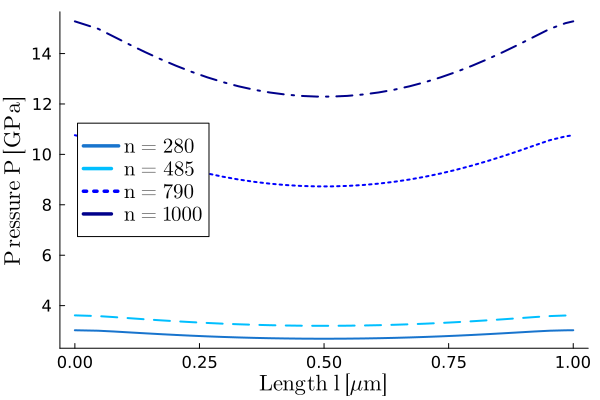}
      (b)
   \end{minipage}
   \begin{minipage}[b]{.45\linewidth} 
   \centering
      \includegraphics[width=\linewidth]{
      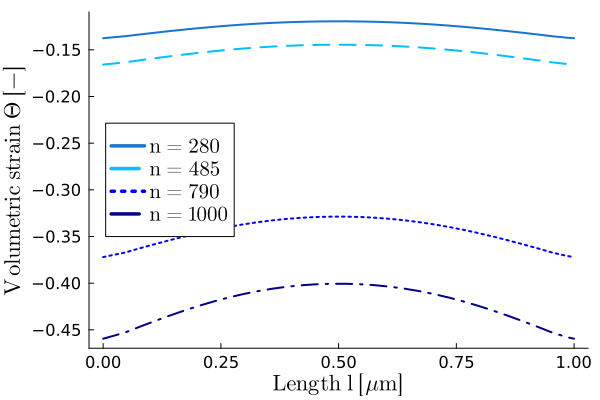}
      (c) 
   \end{minipage}
 \begin{minipage}[b]{.45\linewidth} 
   \centering
      \includegraphics[width=\linewidth]{
      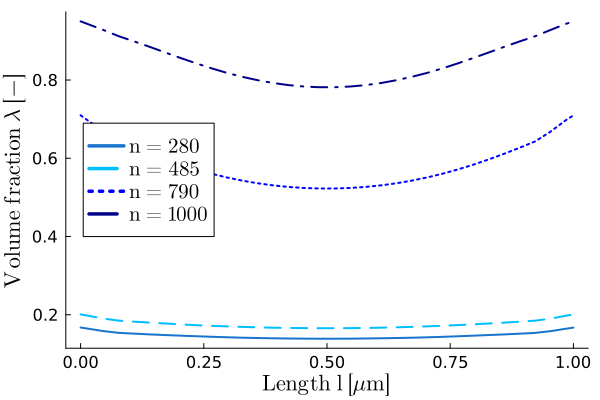}
      (d) 
   \end{minipage}
   \caption{ Time evolution over position x at y = h/2 for the rectangular plate during an uniaxial compression test, for the time steps n = [280, 485, 790, 1000]. (a) shows the development of bulk modulus K, (b) the pressure behavior, (c) volumetric strain $\Theta$ and (d) increasing of volume fraction $\lambda$. A visual representation of phase deveolpment can be seen in Fig.~\ref{fig:12}.\raggedright }
      \label{fig:13}
\end{figure}
\begin{table}[h]
\centering
\caption{ Bulk modulus values for 2d uniaxial compression, coupled shear-compression test and quaterplate with a circular hole. }\label{tab5}%
\begin{tabular}{@{} c | c | c | c   @{}}
\hline

$\mathrm{K_0}$ [GPa] & $\mathrm{K_1}$ [GPa]& $\mathrm{K_2}$ [GPa]& $\mathrm{K_3}$ [GPa]\\
\hline

 30.0 &36.8  & 61.1& 209.0 \\

\hline
\end{tabular}
\end{table}
\begin{table}[h]
\centering
\caption{ Parameters for 2d uniaxial compression, coupled shear-compression test and quaterplate with a circular hole.  }\label{tab5.1}%
\begin{tabular}{@{} c | c | c | c | c  | c   @{}}
\hline

$r_1$ [GPa]&$a$ [GPa] & $\Theta_{\mathrm{T}}$ [-]& $\Theta_{\mathrm{a}}$ [-] & $\eta$ [Pa]&$c$ [GPa]\\
\hline

1.3& 0.47 &21.0   & 0.087& 0.5 $\cdot 10^{9}$&1.1 \\

\hline
\end{tabular}
\end{table}
\begin{figure}[tp]
\centering
  \begin{minipage}[b]{.4\linewidth} 
   \centering  
  \centering
      \includegraphics[width=\linewidth]{
     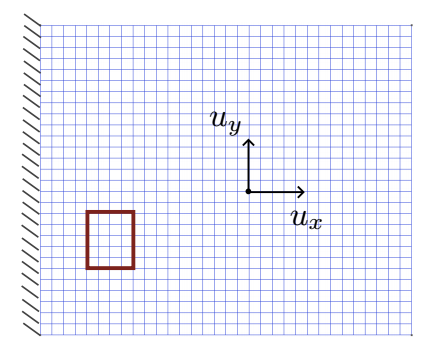}
   \end{minipage}
 \caption{Illustration of boundary conditions and displacements during a coupled shear - compression test. Some elements are marked for data extraction, see Fig. \ref{fig:17}. \raggedright }
      \label{fig:15}
\end{figure}
\subsubsection{ Coupled shear and compression test}\label{subsubs3}
Figure \ref{fig:15} shows the boundary conditions for a coupled shear - compression test. Contrary to the previous example, the displacements in x and y directions are set in a specific ratio, to better understand the non-hydrostatic effects and the interaction between shear effects with plasticity. Therefore, the same geometry and material parameters are used as in the uniaxial compression test, see Fig.~\ref{fig:11} and Tab.~\ref{tab5}, \ref{tab5.1}. Likewise as above, $\mathrm{n_{max}}$ = 1400 time steps are performed, with an animation time of t = 2 $\si{\micro\second}$. 
The boundary conditions are,
\begin{align*}
u_x &= 0~\mathrm{for}~x = 0\\
  u_x &= -\tilde{u}_x(t)~\mathrm{for}~x = l \\
 u_y &= 0~\mathrm{for}~x = 0\\
  u_y &= -\tilde{u}_y(t)~\mathrm{for}~x = l \hspace{0.1cm}.
\end{align*}

\begin{figure}[tp]
\centering
  \begin{minipage}[b]{.9\linewidth} 
   \centering
       \includegraphics[width=\linewidth]{
     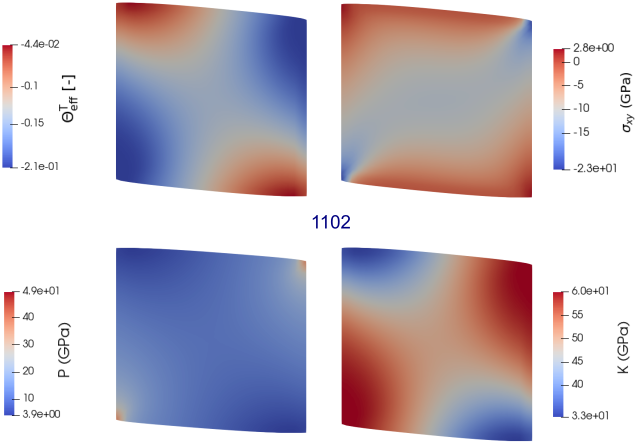}
   \end{minipage}
 \caption{Contour plots of a coupled shear-compression test at time step number n = 1102. From left to right, effective plastic strain $\Theta^{\mathrm{T}}_{\mathrm{eff}}$, deviatoric stress $\sigma_{xy}$, hydrostatic pressure P and bulk modulus K. To demonstrate the deformation, a factor of k = 0.3 is applied.  \raggedright }
      \label{fig:16}
\end{figure}
\begin{figure}[h]
\centering
   \begin{minipage}[b]{.45\linewidth} 
   \centering
      \includegraphics[width=\linewidth]{
      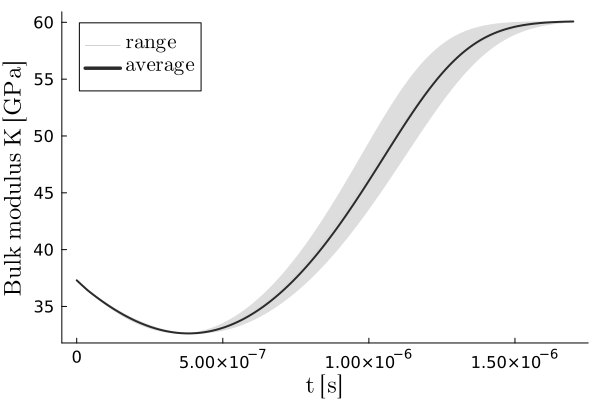}
      (a) 
   \end{minipage}
  \begin{minipage}[b]{.45\linewidth} 
   \centering
      \includegraphics[width=\linewidth]{
      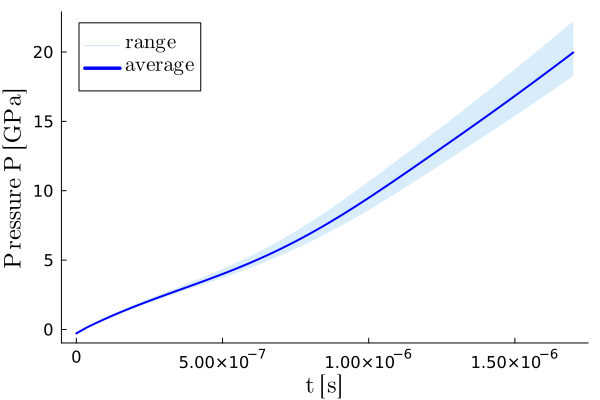}
      (b)
   \end{minipage}
   \begin{minipage}[b]{.45\linewidth} 
   \centering
      \includegraphics[width=\linewidth]{
      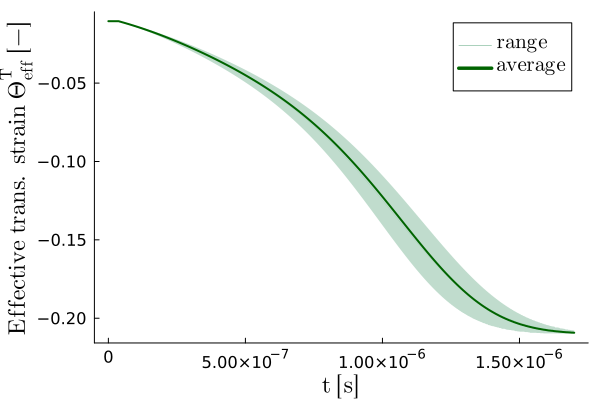}
      (c) 
   \end{minipage}
 \begin{minipage}[b]{.45\linewidth} 
   \centering
      \includegraphics[width=\linewidth]{
      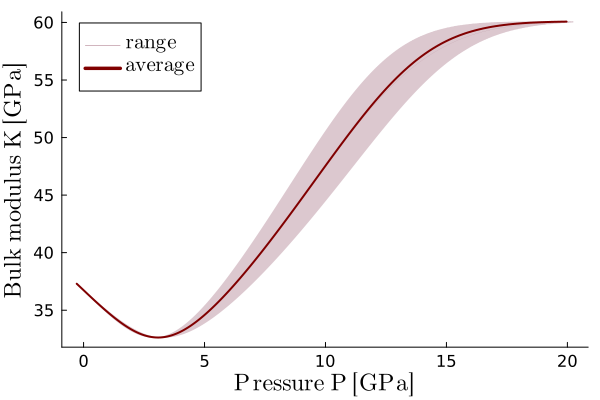}
      (d) 
   \end{minipage}
 
   \caption{From (a) to (c), time development of average element data with range of maximum to minimum values of bulk modulus K,  hydrostatic pressure P and effective plastic strain $\Theta^{\mathrm{T}}_{\mathrm{eff}}$ over some elements during the shear-compression test. The elements are marked with a red rectangle in Fig.~\ref{fig:15}. (d) Bulk modulus K against hydrostatic Pressure P.\raggedright }
      \label{fig:17}
\end{figure}
The ratio of the specified displacement in x to y direction amounts to 1.26 [-]. During deformation the maximum possible hydrostatic pressure is reached at the bottom left and top right corners, as can be seen in the overview at time step number n = 1102 in Fig.~\ref{fig:16}. Similarly, densification starts at the lower left and upper right corner progressing to the center of the sample. Figure \ref{fig:17} shows the average value of several elements, compared with their range of minimum and maximum values. The start and end of the phase transition evolving in the chosen elements appear to coincide with the same values. These ranges could be related to the elastic behavior. It is assumed that the elastic displacements have a lesser influence on the individual Gaussian points. Furthermore, for this example, permanent densification starts at a pressure between (1-2) GPa at approx. t = $5 \cdot 10^{-8}$ \si{\second}, see also Fig~\ref{fig:17} (c). In the region of the suspected phase transition, the smallest and largest element values diverge significantly. This is where the influence of shear effects becomes apparent. \\
On the other hand, Fig. \ref{fig:17} (b) shows an expansion of element data with increasing hydrostatic pressure.
\begin{figure}[h]
\centering
  \begin{minipage}[b]{.85\linewidth} 
   \centering
       \includegraphics[width=\linewidth]{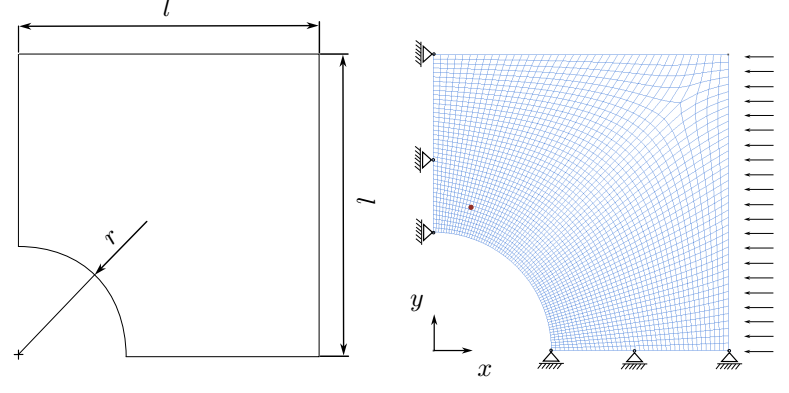}
   \end{minipage}
 \caption{Geometric illustration of the two-dimensional square plate with a circular hole on x - y plane. Due to symmetry, only a quater of the system is modeled. Dimensions on the left side with l = 0.25 mm and r = 0.9 mm, boundary conditions, compressive displacement $u_x$ and quadrilateral mesh with 3016 elements on the right. One element is marked with a red dot. \raggedright }
      \label{fig:18}
\end{figure}
\subsubsection{ Quadratic plate with a circular hole}\label{subsubs4}

\begin{figure}[htp]
\centering
   \begin{minipage}[b]{1.0\linewidth} 
   \centering
      \includegraphics[width=\linewidth]{
    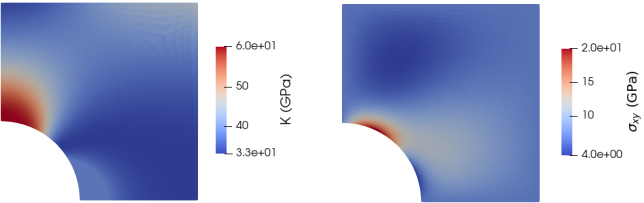}
   \end{minipage}
   \caption{Contour plot of bulk modulus K and deviatoric stress tensor $\boldsymbol{\sigma}_{xy}$ at maximum time step n = 1499 for the example in Fig.~\ref{fig:18}.\raggedright }
      \label{fig:19}
\end{figure}
\begin{figure}[tp]
\centering
  \begin{minipage}[b]{.9\linewidth} 
   \centering
       \includegraphics[width=\linewidth]{
     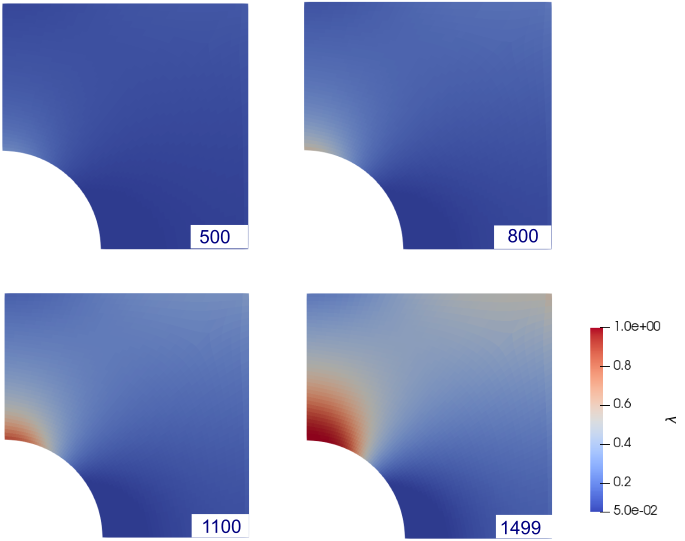}
   \end{minipage}
 \caption{Starting of phase development as contour plot over four time steps  n = [500, 800, 1100, 1499]. Phase evolution begins at the upper region of the hole. \raggedright }
      \label{fig:20}
\end{figure}
Finally as benchmark problem, a quadratic plate with a circular hole is illustrated in Figure \ref{fig:18}. Due to the symmetry only a quater of the domain is modeled. The associated boundary conditions are:
\begin{align*}
u_x &= 0,~\mathrm{for}~ y = 0  \\ u_y &= 0,~\mathrm{for}~ x = 0 .
\end{align*}
\begin{figure}[h]
\centering
   \begin{minipage}[b]{.45\linewidth} 
   \centering
      \includegraphics[width=\linewidth]{
      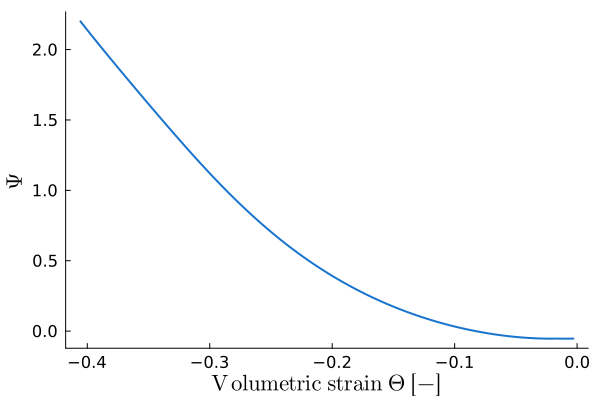}
      (a) 
   \end{minipage}
  \begin{minipage}[b]{.45\linewidth} 
   \centering
      \includegraphics[width=\linewidth]{
   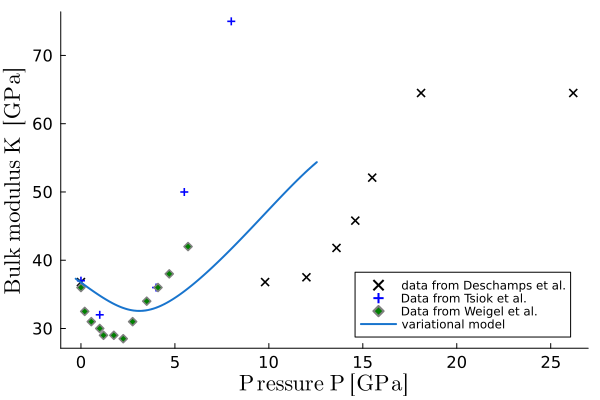}
      (b)
   \end{minipage}
   \begin{minipage}[b]{.45\linewidth} 
   \centering
      \includegraphics[width=\linewidth]{
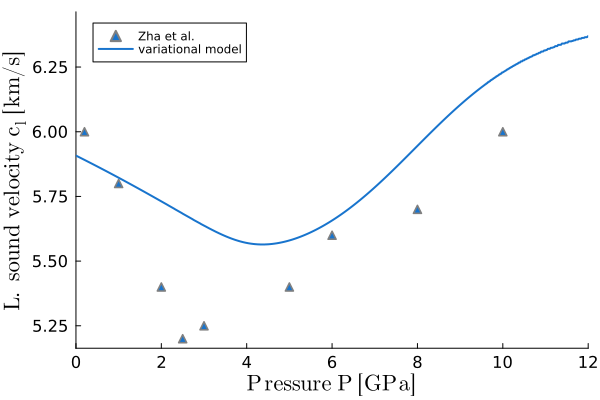}
      (c) 
   \end{minipage}
 \begin{minipage}[b]{.45\linewidth} 
   \centering
      \includegraphics[width=\linewidth]{
     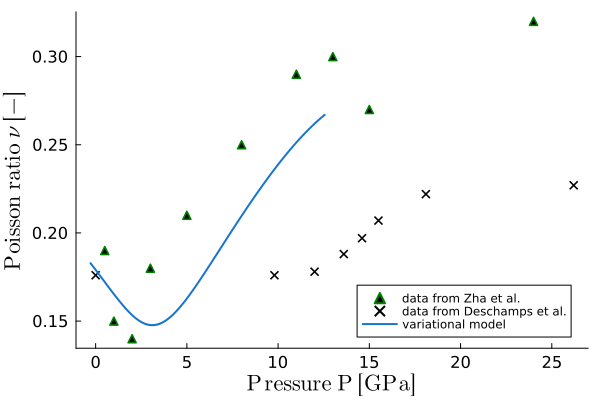}
      (d) 
   \end{minipage}
   \caption{Results of a plate with circular hole at one Gaussian point, showing graphs together with the relaxed energy potential $\Psi$. (a) Relaxed energy potential against volumetric strain $\Theta$, (b) Bulk modulus as function of hydrostatic pressure P with experimental data from \cite{Deschamps14,Tsiok,Weigel}, (c) longitudinal sound velocity $\mathrm{c_l}$ against hydrostatic Pressure P with data from \cite{Zha}. (d) Poisson ratio $\nu$ as function of hydrostatic pressure P with experimental data taken from \cite{Deschamps14,Zha}.  \raggedright }
      \label{fig:21}
\end{figure}
The plate with length $l$ = 0.25 mm and radius $r$ = 0.9 mm is subject to a negative displacement in x-direction on the right hand side. A fine mesh with 3016 elements is used now. Contour plots in Fig.~\ref{fig:19} and Fig.~\ref{fig:20} depict the starting of phase evolution at the boundary of the hole, the associated development of bulk modulus K as well as the calculated deviatoric stress tensor $\sigma_{xy}$ at time step n = 1499.
Across the plate the bulk modulus shows a similar behavior as the volume fraction development. The characteristic minimum of the bulk modulus occurs between the lower edge toward the hole.\\
In summary we register, that these representations show behavior comparable to the examples shown previously. In conclusion, Fig.~\ref{fig:21}  illustrates results at one Gaussian point from an element marked with a red dot, see Fig.~\ref{fig:18}, together with the associated relaxed energy $\Psi$ to verify its convexity for the applied parameters listed in Tab.~\ref{tab5} and \ref{tab5.1}. The material point behavior is presented together with experimental data from \cite{Zha,Deschamps14,Tsiok, Weigel}. The data obtained from the sources mentioned above serve to provide an overview of the progression of the curves. The figures do not contain any error values. The Poisson ratio starts at $\nu \approx$ 0.18, which is in good agreement with experimental data. Otherwise it shows a clear minimum before it increases to $\nu \approx$ 0.3. On the one hand, this trend contrasts with \textit{ex situ} measurements of Deschamps et al. \cite{Deschamps14}, but on the other hand, it shows similarities with \textit{in situ} evaluations of Zha et al. \cite{Zha}. At this point, it should be mentioned that in \cite{Deschamps14} the sensitivity of the measurement method, \textit{in situ} versus \textit{ex situ}, to elastic behavior and the region of transformation is discussed.  This can be confirmed in this context, insofar as a connection to the other graphs is also apparent here. If an anomalous behavior becomes visible, which is part of the reversible region \cite{Mysen}, this is also reflected in the other curves. To confirm this, Fig. \ref{fig:21} (c) shows the longitudinal sound velocity $\mathrm{c_l}$ of our variational model. Results are evaluated using Eq. \eqref{eq:17.1} with an initial density of $\rho_0 = 2.20~ \si{\g\per\cm^3}$ and $\Theta^{\mathrm{T}}_{\mathrm{eff}}$.

\section{Conclusion and Outlook}\label{sec12}

In this work, we present a variational model for the volumetric behavior of \ce{SiO2} glass. The model based on the variational description of inelastic materials is  formulated in strain space. We assume that the sigmoidal progression of the often used equation of state (EOS), representing relative density change as function of pressure, is due to a macroscopic phase transition between two phase fractions. For this behavior, we include an energy density term of ideal binary phase mixing into the model. Moreover, the dissipation regarding the compaction process is formulated employing the so-called dissipation distance to initialize a instantaneous energy loss during the transition process. The rate of the evolving phase fraction gives then access to the phase development. \\
Our numerical results correspond closely to the well-known graphs of silica glass obtained from static measurements. This includes in one hand the anomalous behavior of the bulk modulus. On the other hand, the direct connection to the longitudinal sound velocity is well represented. \\ 
The sigmoidal phase evolution analyzed here can also be found in other glasses with normal or intermediate behavior regarding the elastic moduli, like for example window glass \cite{Rouxel}. The modeled transformation of volume fractions is valid for arbitrary stiffness properties of the solid, hence it can be used to describe other glasses under the same conditions. \\ 
Although this variational model describes the characteristics of silica glass under compression and its resulting volumetric densification, shear effects, which cannot be neglected in amorphous solids, were not be taken into account. However, Finite Element simulations along with employing Hooke's law provide a way to capture Poisson's ratio and give an insight into material behavior under non-hydrostatic boundary conditions. Moreover, there is a lack of a more detailed description of the underlying irreversible transformation in the short-range order. In addition, the model only reflects isothermal situations. Finally, the model in its present form is already able to capture unloading behavior. However, more work on the constitutive equations will be needed to represent experimental results in a satisfactory manner.
\section*{Acknowledgements}

This work has been founded by the Deutsche Forschungsgesellschaft (DFG). We gratefully acknowledge the support within the projekt SPP 2256 ("Variational Methods for Predicting Complex Phenomena in Engineering Structures and Materials"), projekt ID 442307685. \\
Also we thank the Julia community.

\bibliographystyle{elsarticle-num-names} 
\bibliography{sn-bibliography,biblio1}






\end{document}